\documentclass[prl,floats,aps,superscriptaddress,showpacs]{revtex4}
\usepackage{amsmath}
\usepackage{graphicx}
\newcommand{\be}{\begin{equation}}
\newcommand{\ee}{\end{equation}}
\newcommand{\ba}{\begin{eqnarray}}
\newcommand{\ea}{\end{eqnarray}}
\newcommand{\nsigma}{\mbox{\boldmath $\sigma$}}
\newcommand{\fex}{\frac{\mbox{e}^{-T_s/T^{\ast}}}{\tau_s} }
\newcommand{\br}{{\bf r}}
\newcommand{\gam}{{\tau_b^{-1}}}
\newcommand{\bv}{{\bf v}}

\begin{document}
\date{\today}
\title{{\bf Dynamics of vibro-fluidized granular gases in periodic structures}}

\author{U. Marini Bettolo Marconi}
\affiliation{Dipartimento di Fisica, Universit\`a di Camerino and
Istituto Nazionale di Fisica della Materia,
Via Madonna delle Carceri, 62032 , Camerino, Italy}
\author{M.Conti}
\affiliation{Dipartimento di Fisica, Universit\`a di Camerino and
Istituto Nazionale di Fisica della Materia,
Via Madonna delle Carceri, 62032 , Camerino, Italy}

%%%%%%%%%%%%%%%%%%%%%%%%%%%%%%%%%%%%%%%%%%%%%%%%%%%%%%%%%%%%%%%%%%%

\begin{abstract}
The behavior of a driven granular gas in a container consisting
of $M$ connected compartments is studied employing a
microscopic kinetic model. After obtaining the governing equations
for the occupation numbers and the granular temperatures of each compartment
we consider the various dynamical regimes.
The system displays interesting analogies with the ordering
processes of phase separating mixtures quenched below the
their critical point. In particular, we show that
below a certain value of the driving intensity the 
populations of the various compartments become unequal and the system forms
clusters. Such a phenomenon is not instantaneous, but 
is characterized by a time scale, $\tau$, which follows a 
Vogel-Vulcher exponential behavior.
On the other hand, the reverse phenomenon which involves the
``evaporation'' of a cluster due to the driving force is also
characterized by a second time scale which diverges at the limit of 
stability of the cluster.

\end{abstract}
\maketitle

%%%%%%%%%%%%%%%%%%%%%%%%%%%%%%%%%%%%%%%%%%%%%%%%%%ti%%%%%%%%%%%%%%%%%
%%%%%%%%%%%%%%%%%%%%%%%%%%%%%%%%%%%%%%%%%%%%%%%%%%%%%%%%%%%%%%%%%%%
%\tightenlines

\section{I. Introduction}

One of the most ubiquitous phenomena in nature is the existence
of matter in different phases. Water, for instance, can form
either a solid, a liquid or a gas. When such a system is brought, by 
a sudden change of a control
parameter, from a one phase equilibrium state , 
to a point in the phase diagram where it can exist in two
different phases, it will reach its new thermodynamic equilibrium
only through a non equilibrium process, called phase separation
\cite{transition}.
In the present paper, we stress that granular fluids
possess a phenomenology which recalls closely that of standard
matter. More specifically, we show that some processes
observed during an ordinary condensation-evaporation process
have a counterpart in  vibrated granular fluids. In particular, 
the formation of dense patches
surrounded by rarefied regions is similar 
to the phase separation dynamics associated
with a liquid-gas transition \cite{goldhirsch}.

The idea that granular fluids \cite{gases}-\cite{Kadanoff}, 
i.e. large collections
of inelastic particles fluidized by the action of an external driving
force, 
under appropriate conditions
may exhibit behaviors typical of ordinary fluids is contained in some
recent papers.  
Sunthar and Kumaran reported the coexistence of
different states in 
vibro-fluidized granular beds \cite{Kumaran}. 
Argentina et al., instead, claimed that
a vibrated gas of inelastic particles displays
Van der Waals loops \cite{Argentina} in a
pressure-density diagram.

Some years ago, Schlicting 
and Nordmeier in a seminal paper \cite{experiment} 
considered an assembly of 
steel balls in a vertical container of height $L$ partitioned in two 
connected sections, by a dividing wall of height $l<L$.
They observed that, when the container
was vigorously shaken, the number of balls in the
two sections was statistically identical, whereas the two
populations were dramatically different for weak shaking.
The reason for such a behavior is the competition
between the particle diffusion induced by the shaking 
and the tendency to cluster resulting from the inelastic collisions.
Various authors contributed with theoretical explanations
of such a problem. These range from phenomenological flux models 
\cite{Eggers}-\cite{Lohse2}, to urn models \cite{Droz}-\cite{Yee}, 
to kinetic theory approaches\cite{Brey},
\cite{ultimonostro},\cite{dererumpallettarum} .
Moreover, the Twente group \cite{Lohse2} presented new 
sets of experiments
which have stimulated more interest in the problem.
%In particular they have considered situations where
%the number of communicating compartments was much larger than two
%and found interesting differences with the case where
%only two compartments are present.

We shall consider a granular gas
subjected to a vigorous shaking and 
initially equi-partitioned into several identical compartments 
and show that it presents a phenomenology 
resembling that of spinodal decomposition 
\cite{spindecomp}. 
At some instant $t=0$ the shaking intensity is decreased and the
system evolves towards a new statistically steady state. The system 
has two possibilities: one is to 
persist in the homogeneous state, the second to
cluster. The
crossover between the two regimes
occurs at a particular value of the driving intensity
through the amplification of long-wavelength fluctuations.
However, the analogy with spinodal decomposition 
is not complete, since in the late stage
the order parameter does not saturate and the behavior of the
granular system shows substantial differences with respect
to familiar ordering systems.

The remainder of the paper is organized as follows. In section II
we specify the model, which is based on the assumption that 
the grains are described as inelastic hard disks subjected to 
a stochastic driving force. It is then possible to derive
within the framework of the Boltzmann equation
the governing law for the occupation number and the
granular temperature of each compartment, the relevant variables 
in the problem. The governing equations represent an extension
of those recently employed in ref.\cite{ultimonostro} and validated
by Monte Carlo simulation.
In section III we perform the stability analysis of the homogeneous
asymptotic solutions of
the governing equations and extract predictions
about the crossover from the homogeneous
regime to the clustered regime. In section IV we illustrate
the numerical results for one and two dimensional systems.
Finally, in section V we present our conclusions.

%%%%%%%%%%%%%%%%%%%%%%%%%%%%%%%%%%%%%%%%%%%%%%%%%%%%%%%%%%%%%%%%%%%%%%%%%5

\section{II. model}
We propose a simple extension of the model employed elsewhere
\cite{ultimonostro}
in order to study the steady state properties of a vibro-fluidized
granular gas to the case where the total volume available to 
the particles is divided into a series of identical compartments
and the grains can move from one to the other by jumping 
over a vertical wall.
An assembly
of ${\cal N}$ inelastic hard-spheres moves in a $d$-dimensional 
domain partitioned
into $M$ identical regions of
volume, $V$ separated by vertical obstacles. 
Each compartment contains $N_i$ particles
so that $\sum_{i=1}^M N_i={\cal N}$. 

When two particles collide their
velocities
after the collision, denoted with a prime, are obtained in terms
of the (unprimed) precollisional velocities through the relation 
\begin{eqnarray}
{\bf v}^{\prime}_1={\bf
v}_1-\textstyle{\frac{1}{2}}(1+\alpha)({\bf v}_1-{\bf v}_2)
\cdot\hat{\nsigma})\hat{\nsigma}
\label{eq:uno}
\end{eqnarray}
where $\hat{\nsigma}$
is the unit vector directed from particle $1$ to particle $2$,  
and $\alpha$ 
is the coefficient of restitution.
A particle in the $i$-th box, besides
colliding inelastically with remaining $(N_i-1)$ within the same
box is subjected to the action of a 
white noise random force,
which compensates the energy losses due to dissipative forces.

The dynamics of the $k$-th
particle between two successive collisions is based
on a Langevin type equation of motion for each grain with a
fluctuating random force accounting for the action of
the external driving force:
\begin{equation}
\frac{{\rm d}{\bf v}_k}{{\rm d} t}= -\frac{1}{\tau_b} {\bf v}_k+ 
{\bf{\xi}_k},
\label{eq:due}
\end{equation}

where $-\gam {\bf v}_k$ is a viscous term and ${\bf {\xi}_k}$ a
Gaussian random acceleration, whose average is zero and variance 
satisfies a fluctuation-dissipation relation:
\begin{equation}
\langle {\xi}_{k\mu}(t) {\xi}_{m\nu}(t^\prime)\rangle =
2 \frac{T_b  }{m \tau_b} \delta_{km}
\delta_{\mu\nu}
\delta(t-t^\prime),
\end{equation}
where $T_b$ is proportional to the intensity of the driving~\cite{puglisi}
and $\mu,\nu$ denote vector components. 
The rate at which the kinetic energy is dissipated
by collisions is proportional to
$1-\alpha^2$.

Finally, the particles contained in 
compartment $i$ can migrate
into compartments $i \pm 1$ with a probability per unit time,
$\tau_s^{-1}$, provided their kinetic energy exceeds the 
fixed threshold $T_s$, which is related to the gravitational
energy necessary to overcome the vertical barrier.

Instead of considering the individual trajectories
of the particles, one can study 
the single particle phase-space
distribution function  $f(\br,\bv,t)$, which contains 
most of the relevant
statistical information about the system. 
In order to obtain $f(\br,\bv,t)$ one still has to solve the 
associated Boltzmann equation for $f(\br,\bv,t)$ \cite{Huang},
which is not a simple task.
However,
a simpler description can be achieved
by assuming that the relevant properties 
of the system are described in terms of the
average particle population and of the average kinetic energy
in each compartment.
Therefore, we introduce the
following coarse grained distributions:
$$
f_i(\bv,t)=\frac{1}{V}
\int_{V_i}{\rm d} {\bf r} f({\bf r},
{ \bf v},t) 
$$
where the integration domain is restricted to the volume 
of the $i$-th compartment.
Such an approximation clearly neglects the gradients which are present
within each box.

Interestingly, also 
the coarse grained distributions $f_i(\bv,t)$ can be obtained
by means of a Boltzmann like kinetic
approach. In fact, $f_i(\bv,t)$
evolves in time due to: a)  the interaction with the heat bath
b) the collisions between the particles
belonging to the same box, c) particle diffusion from one compartment
to the other.
The effect due to a) is represented
by a Fokker-Planck term:
\be
\frac{1}{\tau_b}\frac{\partial}{\partial{\bf v}_1}(\frac{T_b}{m} 
\frac{\partial}{\partial{\bf v}_1}
+{\bf v_1}) f_i(\bv_1,t)
\ee
whereas the effect of b) is encapsulated in the collision term,
$I(f_i,f_i)$ (see ref. \cite{vannoije}).
Finally, we add a term  
taking into account the flux of particles between 
neighboring compartments. Since only fast particles
contribute to the latter process such a term contains 
a Heaviside $\theta$ function.
Collecting terms
we obtain the following equation of motion for $f_i$
\be
\partial_t f_i(\bv_1,t)=I(f_i,f_i)+\frac{T_b}{ m \tau_b} 
\left(\frac{\partial}{\partial{\bf v}_1}\right)^2 f_i(\bv_1,t)+
\frac{1}{\tau_b} \frac{\partial}{\partial{\bf v}_1} ({\bf v}_1 f_i(\bv_1,t))-
\frac{1}{\tau_s}\theta(|\bv_1|-u_s)
[2 f_i(\bv_1,t)-f_{i+1}(\bv_1,t)-f_{i-1}(\bv_1,t)]
\label{eq:boltzmann}
\ee
 A more detailed treatment of
such transport equation can be found in reference \cite{puglisi}
together with \cite{ultimonostro}.

In order to characterize the macrostate of the system we only
need the average number, $N_{i}(t)$, of particles
in compartment $i$ at instant $t$, 
and the corresponding granular temperature $T_{i}$.
These quantities are related to $f_i$ by the equations
\be
N_{i}(t)=\int{\rm d}\bv f_i({\bf v},t)
\ee
where  and $V_i$ is the volume of
the same compartment, and
\be
T_{i}(t)=\frac{1}{N_{i}(t) d}
\int {\rm d}\bv m \bv^2
f_i(\bv,t).
\ee

The problem can be reduced to a simple set of governing equations for the
occupation numbers and the temperatures of the various compartment
(see ref. \cite{ultimonostro}  for details)
if one assumes a Gaussian shape for the distribution function
$f_i$ and $d=2$
\be
f_i(\bv,t)=\frac{N_i}{V}\frac{1}
{2 \pi T_i}\exp(-\frac{\bv^2}{2 m T_i}).
\label{gauss}
\ee

In the case of non communicating compartments
($\tau_s \to \infty$), each containing
$N^{\ast}={\cal N}/M$ particles, the granular temperatures
$T_i$ assume the value $T^{\ast}$,
determined by the solution of the equation 
\be
T^{\ast} \left[1+\tau_b\sigma
(1-\alpha^2)\frac{N^{\ast}}{2 V}\sqrt{\frac{T^{\ast}}{m}}\right]=T_b
\label{uniformT}
\ee

In the general case,
after simple manipulations, we obtain the following set 
of coupled equations

\begin{subequations} 
\label{eq:governing}
\begin{align}
\frac{{\rm d} N_i(t)}{{\rm d}t}&=\frac{1}{\tau_s}
\left[ N_{i+1} e^{-T_s/T_{i+1}}+N_{i-1} e^{-T_s/T_{i-1}}
-2 N_i e^{-T_s/T_i}\right]
\label{eq:exchA} \\
\begin{split}
N_i \frac{{\rm d} T_i(t)}{{\rm d}t}&=
\frac{1}{\tau_s}
[ 2(N_{i+1} T_{i+1} e^{-T_s/T_{i+1}}+
N_{i-1} T_{i-1} e^{-T_s/T_{i-1}}
-2 N_i T_i e^{-T_s/T_i})\\
&+(N_{i+1} e^{-T_s/T_{i+1}}+N_{i-1} e^{-T_s/T_{i-1}}
- 2 N_i e^{-T_s/T_i})(2 T_s-T_i)] 
-2 \gamma \omega_i N_i T_i +\frac{2} {\tau_b}N_i( T_b-T_i) 
\label{eq:exchB} 
\end{split}
\end{align}
\end{subequations}

where the local dissipation rate \cite{vannoije} is

\be
\gamma \omega_i=\sigma
(1-\alpha^2)\frac{N_i}{2 V}\sqrt{\frac{T_i}{m}}
\ee
Let us notice that the ansatz (\ref{gauss}) is 
only dictated by technical convenience and not necessary,
since it leads to a small set of coupled equations
for $N_i$ and $T_i$. In fact,
one way to improve systematically the approximation
is to assume $f_i$ to be given
by a Gaussian multiplied by
a linear combination 
of orthogonal polynomials, the so called Sonine polynomials.
However, in this case
one should consider extra variables besides $N_i$ and $T_i$
and the resulting set of equations would be of higher rank.

We turn now to consider the most relevant behaviors associated
with the governing equations (\ref{eq:exchA}) and (\ref{eq:exchB}).
Incidentally we notice that eqs. (\ref{eq:exchA}) and(\ref{eq:exchB})  
in the case of $M=2$ are
identical to the corresponding equations of reference
\cite{ultimonostro} after substituting $\tau_s$ with $2 \tau_s$.

A large collection of grains initially distributed
uniformly in an array of $M$
identical compartments may remain homogeneous or not
according to the intensity of the driving at which is
subjected. The Twente group observed experimentally
a homogeneous
configuration under vigorous shaking, and
clusters for weak shaking \cite{Lohse2}. The crossover from the first
to the second regime occurs when the ``thermally'' induced
diffusion is not sufficient to prevent the spontaneous tendency
of the grains to form clusters, due to the collisional cooling. 
 Such an instability is the counterpart of the separation
process which occurs in a system consisting of two 
compartments only. In such a case the
left-right symmetry, i.e. the
difference between the 
left population, $N_L$,  and the right
population, $N_R$, is spontaneously broken below a certain critical
temperature. Fig. 1a illustrates
the corresponding behavior of the
asymmetry parameter $A=|N_L-N_R|/{\cal N}$ versus the
total number of particles for a fixed value of $T_b=0.7$,
whereas in Fig. 1b we display the variation of
the granular temperatures within the two compartments
versus the total number of particles.
The curve bifurcates at a critical values of the total number
of particles identifying a
temperature, $T_c$, below which the left-right symmetry
is broken.
  We shall discuss how such a mechanism manifests in the case of many
compartments and gives rise to a phenomenology similar to that
of the spinodal decomposition \cite{spindecomp}.
In general
the space of dimensionless control parameter is large since
the system properties are functions of $\alpha$, $M$, $\cal{N}$,
$V/\sigma^2$, $T_b/T_s$, $\tau_b/\tau_s$ and $\gamma\tau_s$.
In the following we shall measure temperatures in units of $T_s$,
areas in units of $\sigma^2$ and time in units of $\tau_b/2$.
%%%%%%%%%%%%%%%%%%%%%%%%%%%%%%% FIG 1 %%%%%%%%%%%%%%%%%%%%%%%%%%%%%%%%%
\begin{figure}[htb]

{\includegraphics[clip=true,width=7.cm, keepaspectratio,angle=0]
{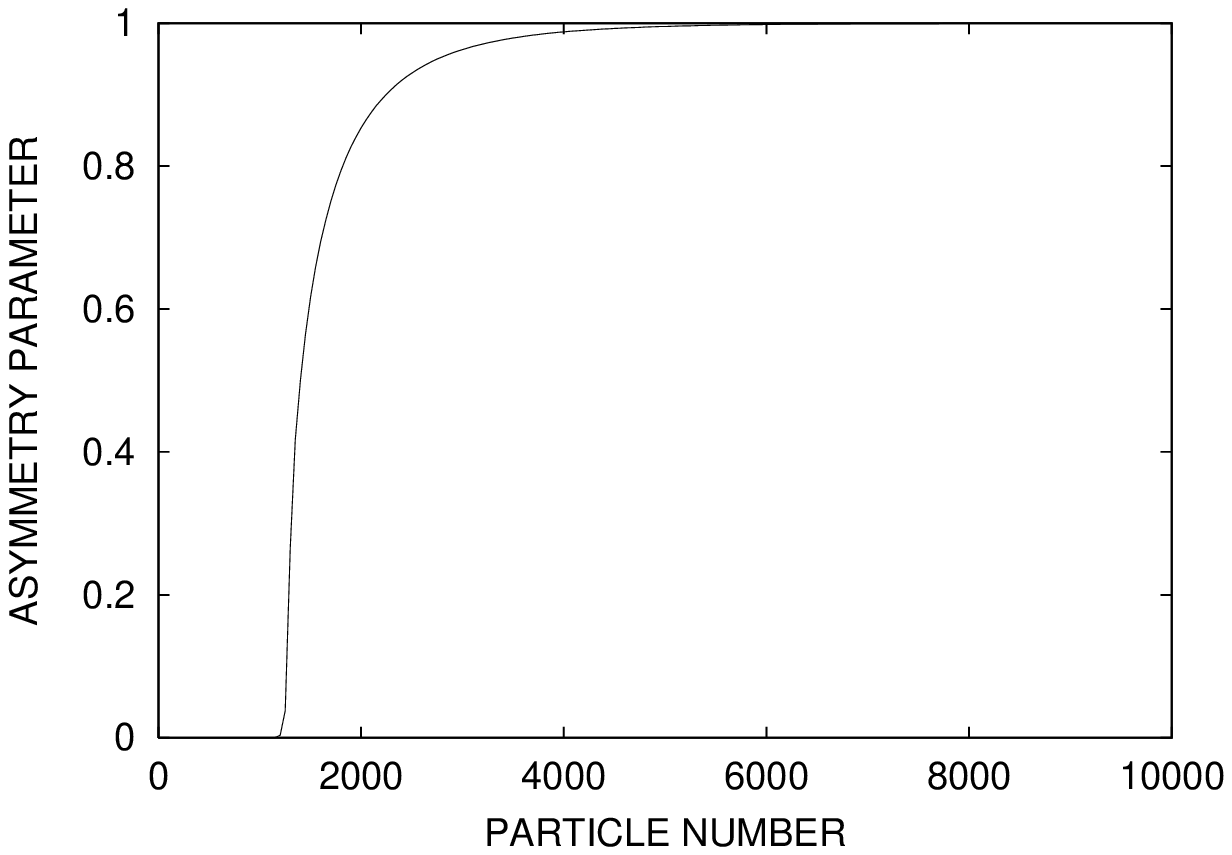}}
\label{Ordpara}
{\includegraphics[clip=true,width=7.cm, keepaspectratio,angle=0]
{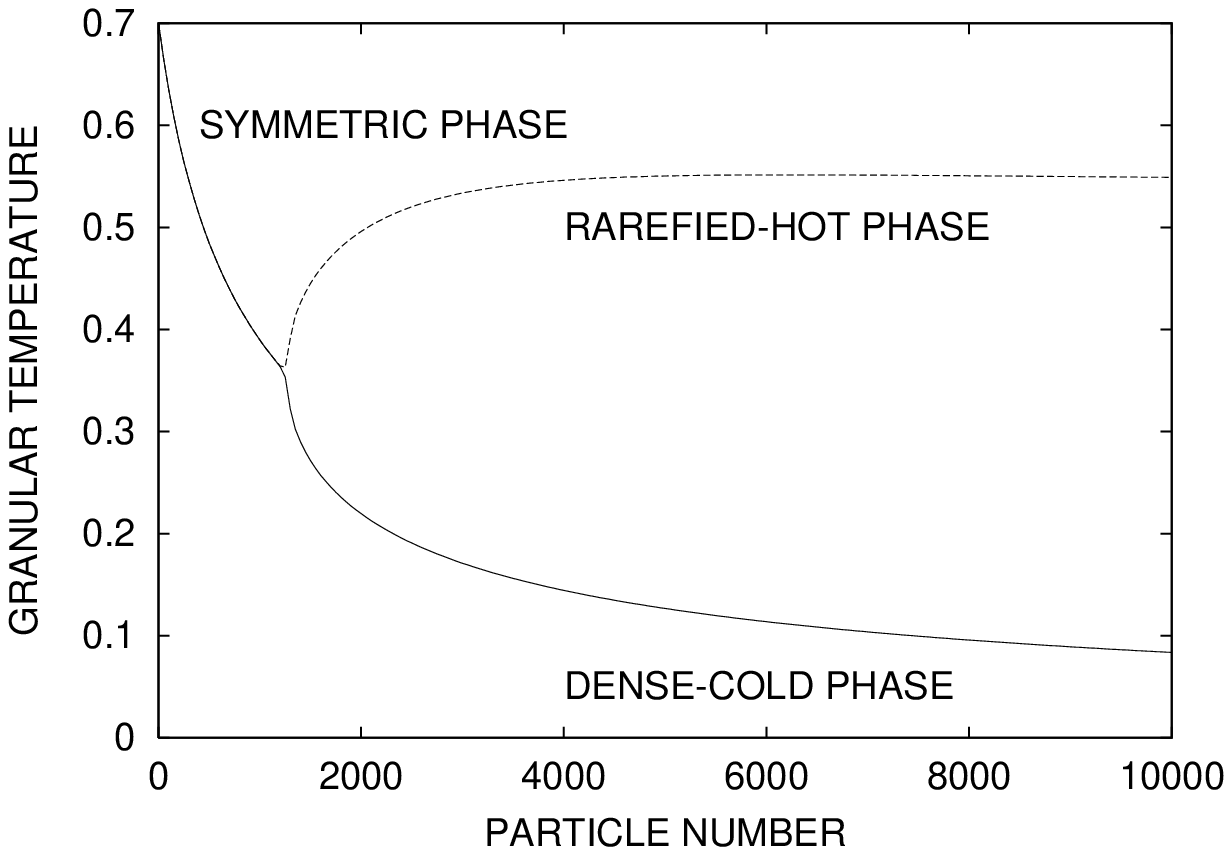}}
\caption
{Phase diagram relative to a system of two 
compartments: in Fig. 1a we show the
asymmetry parameter $A=|N_L-N_R|/{\cal N}$ as
a function of the total number of particles, ${\cal N}$. In Fig. 1b 
granular temperatures of the two compartments are displayed 
as a function of ${\cal N}$.
The heat-bath temperature is fixed at $T_b=0.7$, $T_s=1$, $\tau_b=2$
and $\tau_s=0.5$, while
the compartment 
volume is $100$ $\sigma^2$ and 
$\alpha=0.7$. The granular temperature is measured in non dimensional units. }
\label{Tg_vs_Ntot}
\end{figure}

\section{III. Stability analysis of the homogeneous state}

In the case of $M$ identical compartments with cyclic
boundary conditions, the choice
$N_i=N^{\ast}$ and $T_i=T^{\ast}$, where $T^{\ast}$
and $N^{\ast}$ are related by eq. (\ref{uniformT})
represents a uniform solution of eqs. (\ref{eq:governing}),
for all values of the
control parameters. 
We observe that $T^{\ast}$ is the 
granular temperature of a system of ${\cal N}$ particles,
equally distributed into $M$ compartments of volume $V$ 
and subjected to a heat bath $T_b$.

On the other hand, it turns
out that such a uniform solution is stable only at high temperature,
where a diffusive mechanism tends to restore any small perturbation about
the homogeneous state. On the contrary, the 
uniform state below a certain temperature turns out to be  
unstable with respect to spontaneous fluctuations, 
due to the clustering mechanism induced by inelasticity. 
As shown in Fig. 1b, the granular temperature in the
case of a system with two compartments
takes on two different values when the total population exceeds a threshold
value. 
In the case of many connected compartments 
a related phenomenon occurs. 
In order to illustrate it, we introduce a 
small amplitude sinusoidal perturbation
$T_l=T^{\ast}+\delta T_k \exp(i k l) $,
and $N_l=N^{\ast}+\delta N_k \exp(ikl)$, 
where $k=2 \pi n/M$, with $n=1,..,M-1$
and $l=1,..,M$ denotes the compartment.

Expanding linearly eqs. (\ref{eq:exchA}) and (\ref{eq:exchB}) about
the symmetric fixed point $T^{\ast},N^{\ast}$ one finds the result: 
%%%%%%%%%%%%%%%%%%%%%%%%55

\begin{subequations}
\begin{align}
\delta\dot N_k &=-\fex 2
 (1-\cos(k)) \left[\delta N_k+\frac{ N^{\ast}T_s}
{(T^{\ast})^2}\delta T_k\right]
\label{eq:gov1} \\
\begin{split}
%%%%%%%%%%%%%%%%%%%%%%%%%%%%%%%%%%%%%
\delta\dot T_k &=-\fex \left[\left(2+\frac{T_s}{T^{\ast}}
+2\left(\frac{T_s}{T^{\ast}}\right)^2\right) 
(1-\cos(k))
+\left(3\gamma\omega^{\ast}
+\frac{2}{\tau_b}\right)\right] \delta T_k \\
&-\frac{2}{N^{\ast}}\left[ \fex (T^{\ast}+2 T_s)
(1-\cos(k))+
\gamma\omega^{\ast}T^{\ast}\right] \delta N_k
\label{eq:gov2} 
\end{split}
\end{align}
\end{subequations}

%%%%%%%%%%%%%%% EIGENVALUES 

%%%%%%%%%%%%%%%%%%%%%%%%%%%%%%% FIG 3 %%%%%%%%%%%%%%%%%%%%%%%%%%%%%%%%%
\begin{figure}[htb]
%\centerline{\includegraphics[clip=true,width=\textwidth, keepaspectratio]
\centerline{\includegraphics[clip=true,width=7.cm, keepaspectratio,angle=0]
{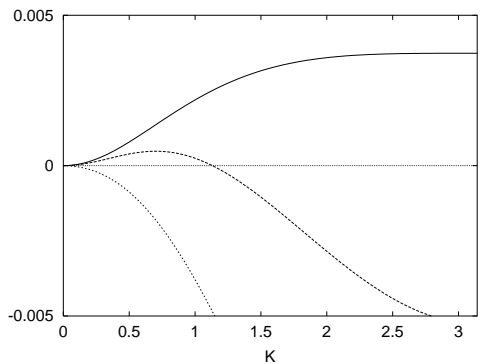}}
\caption
{Variation of the eigenvalue $\lambda_n(k)$ with 
respect to the wave-vector $k$ for different choices of the heat-bath
temperature $T_b=0.28$ (continuous line), $T_b=0.38$ (dashed line) and 
$T_b=0.45$ (dotted line). 
The data refer to a system
consisting of $100$ compartments, each initially
filled with $180$ particles. The compartment 
volume is $100$ $\sigma^2$ and  
$\alpha=0.7$. 
The threshold temperature is $T_s=1$, $\tau_b=2$, $\tau_s=0.5$:
Notice that the mode becomes unstable, i.e.  $\lambda_n(k)$ becomes 
positive, below the critical temperature.}
\label{dispersion}
\end{figure}

%%%%%%%%%%%%%%%%%%%%%%%%%%%%%%% FIG 3 %%%%%%%%%%%%%%%%%%%%%%%%%%%%%%%%%
The associated eigenvalues $\lambda_n(k)$ and 
$\lambda_T(k)$ of the dynamical matrix of coefficients
correspond to the two  relaxation modes
of the system. The larger eigenvalue $\lambda_n(k)$
vanishes quadratically when $k \to 0$,
reflecting the conservation of the global number of particles.
However, due to the coupling between the density and the thermal
fluctuations, for finite $k$, $\lambda_n(k)$ 
presents a variety of behaviors,
associated with different physical phenomena.

In order to classify these behaviors, we  
consider the following expansion valid for small $k$ values
\be
\lambda_n(k)=a_2 k^2+ a_4 k^4 
\label{lambda}
\ee
Above $T_c$, $a_2$ is negative, the eigenvalue
$\lambda_n(k)$ describes a diffusion 
process by which a local density fluctuation is re-adsorbed.
In other words, the particle fluxes caused by the coupling to the 
external driving are sufficient to restore homogeneity.

Below a certain characteristic temperature, $T_c$,
a fluctuation which increases locally the population
is amplified. The local granular temperature
drops due to the increased collision rate,
since (from eq. (\ref{uniformT}))  
$\omega_l \propto N_l T_l^{1/2}
\propto N_l^{2/3}$, and $T_l \propto  N_l^{-2/3}$. 
Thus the particles arriving from the
other compartments remain trapped, causing a further reduction
of the local temperature.
This phenomenon is described by the formula 
\be
a_2=\fex\left[\frac{T_c}{T^{\ast}}-1\right]
\label{a2}
\ee
i.e., $a_2$ becomes positive below $T_c$ which is given by
\be
T_c=\frac{T_s} {\frac{3}{2}+\frac{1}
{\tau_b \gamma\omega^{\ast}}}
\ee
with $\gamma\omega^{\ast}= \sigma
(1-\alpha^2)\frac{N^{\ast}}{2 V}\sqrt{\frac{T^{\ast}}{m}}$.
Contrary to the previous diffusive case, particles
tend to cluster.

According to the sign of $a_4$ (for $a_2>0$) the initial regime 
is different. In fact,
if $a_4<0$ , $\lambda(k)$ may
display a maximum at a finite wave-vector, $k_m<2\pi$
($T_b=0.38$, in fig.~\ref{dispersion}),
whereas for $a_4>0$,  $\lambda_n(k)$ attains its maximum 
only in correspondence of the largest wave-vector for $T_b=0.28$.

Correspondingly, in the second case the growth process
initially resembles the early stage of the spinodal decomposition
process, during which
the homogeneous state is unstable with respect to long wavelength
fluctuations. These are exponentially amplified, whereas short scale
fluctuations decay.

Physically speaking,
a local density increment  induces  
a decrement in the granular temperature. If the
temperature of the bath is sufficiently low
or the number of particles sufficiently large this leads to an instability,
i.e. more particles will flow toward the region where the temperature 
is lower. Diffusion will not be able to compensate such a
tendency.

Finally, there exist a second collective mode $\lambda_T(k)$,
which essentially describes how temperature fluctuations decay.
It is always negative, due to the presence of dissipation caused by friction
and collisions, and given by
\be
\lambda_T(k)=
-\left(3\gamma\omega^{\ast}
+\frac{2}{\tau_b}\right)-c k^2.
\ee

\section{IV. Numerical results}

%%%%%%%%%%%%%%%%%%%%%%%%%%%%%%% FIG 4 %%%%%%%%%%%%%%%%%%%%
\begin{figure}[htb]
{\includegraphics[clip=true,width=5.cm, keepaspectratio,angle=0]
{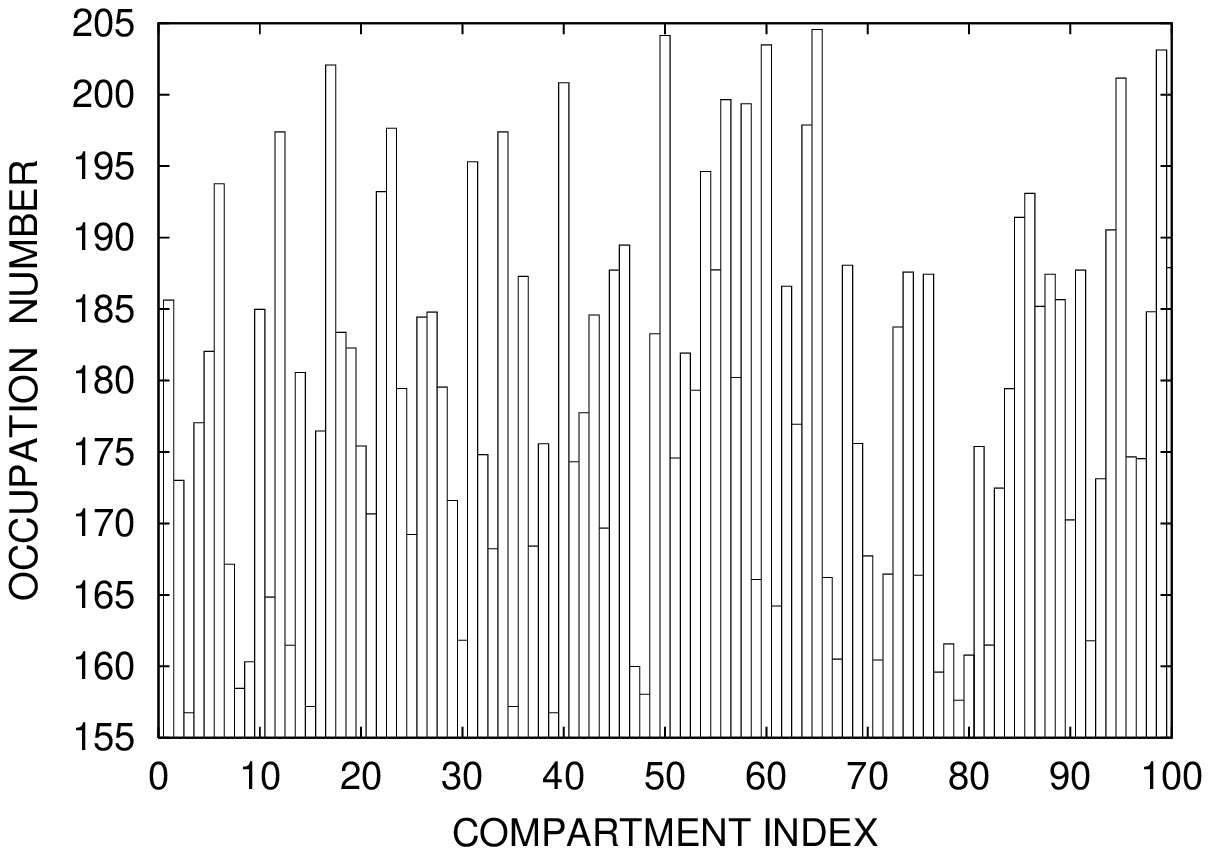}}
{\includegraphics[clip=true,width=5.cm, keepaspectratio,angle=0]
{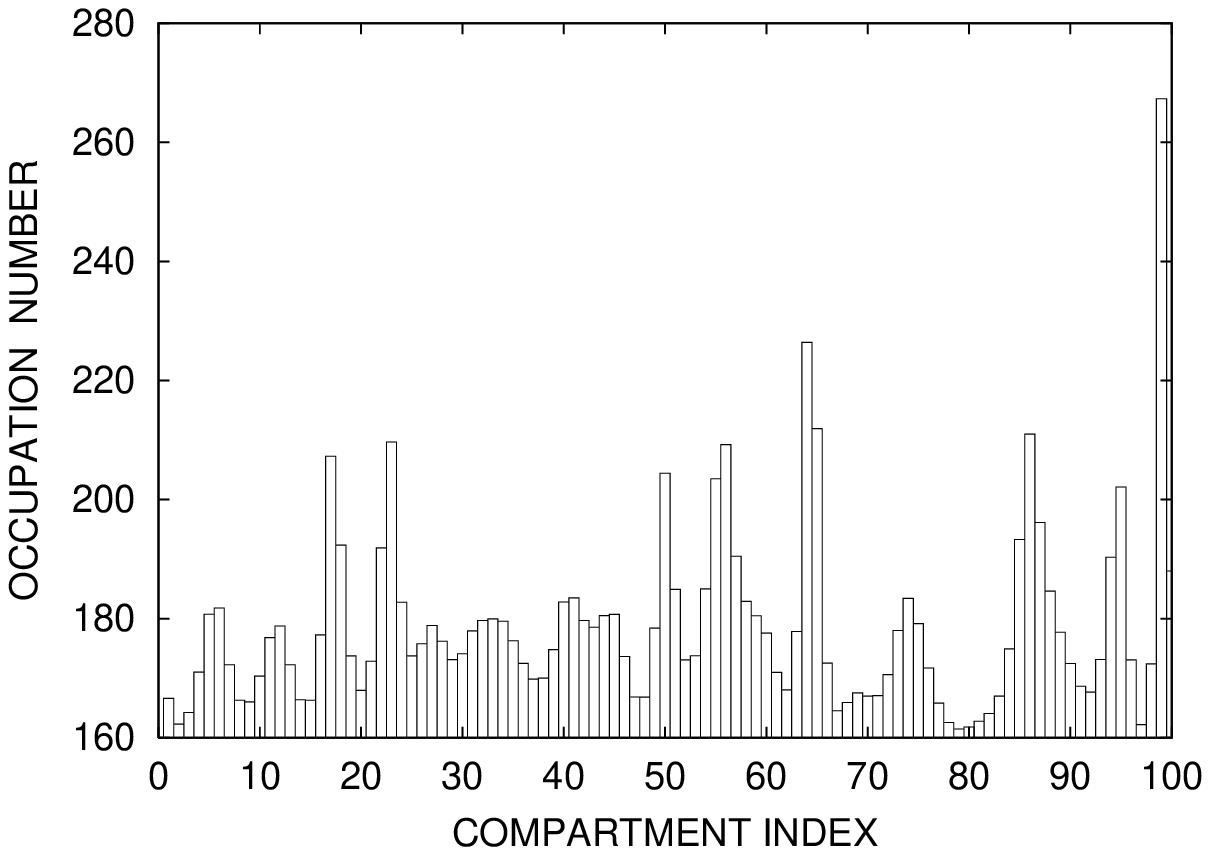}}
{\includegraphics[clip=true,width=5.cm, keepaspectratio,angle=0]
{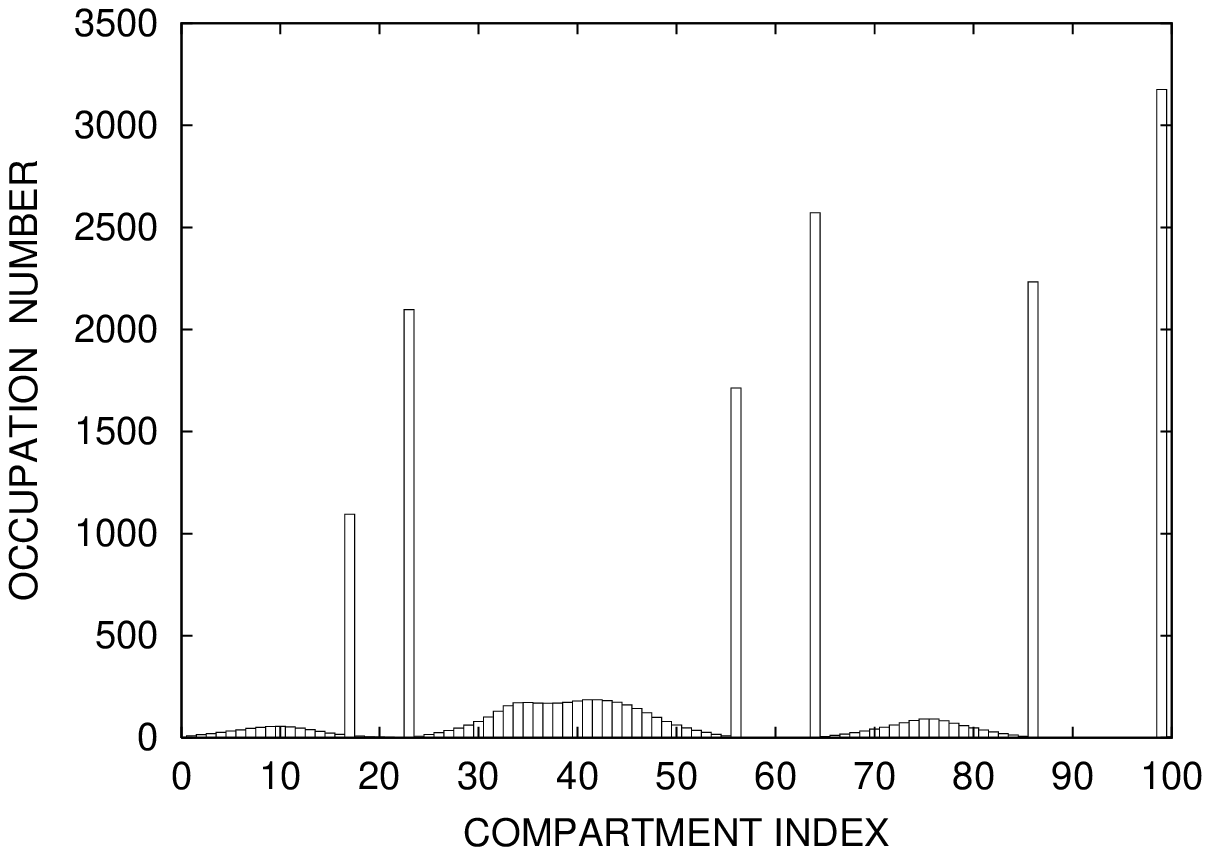}}
\caption
{Evolution of the occupation profile relative to a system 
constituted by $100$ compartments and $180$ particles initially 
placed in each
compartment. The temperature, is chosen below the critical temperature
and is $T_b=0.38$, while the remaining parameters 
are the same as in fig.~\ref{dispersion}. 
The three snapshots from left to right refer to times $t_1=500$,
$t_2=1500$ and $t_3=3000$ in units of $\tau_b$.}
\label{unstable}
\end{figure}
%%%%%%%%%%%%%%%%%%%%%%%%%%%%%%%%%%%%%%%%%%%%%%%%%%%%%%%%%%%%%
%%%%%%%%%%%%%%%%%%%%%%%%%%%%%%% FIG 4added %%%%%%%%%%%%%%%%%%%%
\begin{figure}[htb]
{\includegraphics[clip=true,width=7.cm, keepaspectratio,angle=0]
{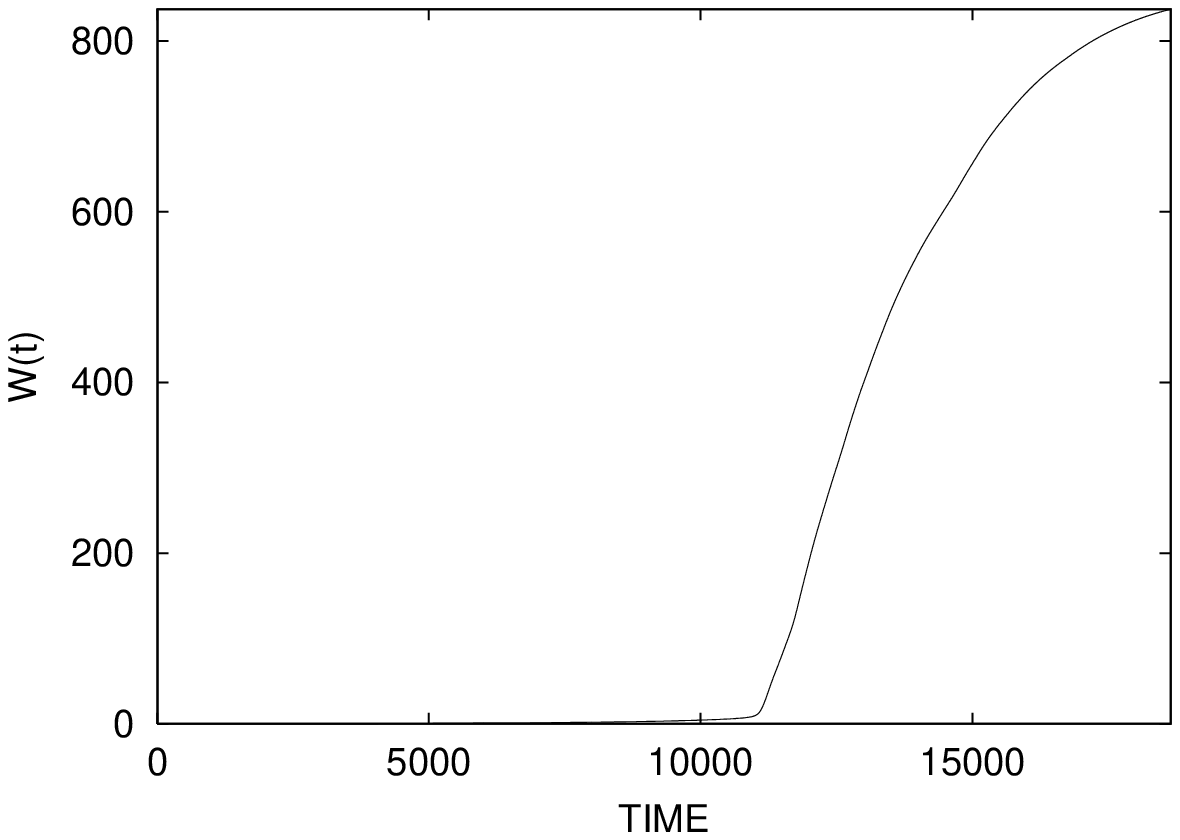}}
{\includegraphics[clip=true,width=7.cm, keepaspectratio,angle=0]
{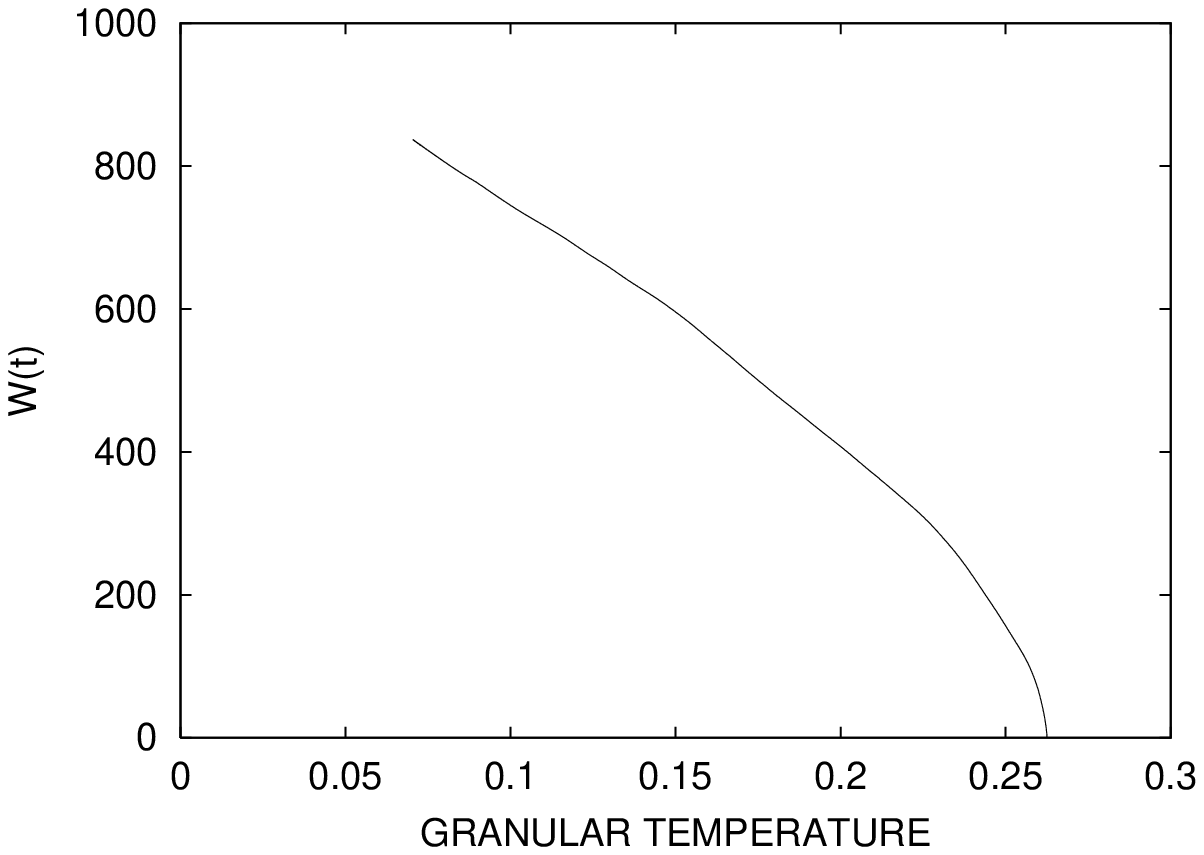}}
\caption
{Growth of the profile width relative to a system 
constituted by $100$ compartments and $180$ particles initially
placed in each
compartment. The heat-bath temperature
is $T_b=0.38$ and the remaining parameters 
are the same as in fig.~\ref{dispersion}.
The right figure represents the parametric plot 
$W(t)$ versus the average granular temperature for the same system.}
\label{added}
\end{figure}
%%%%%%%%%%%%%%%%%%%%%%%%%%%%%%%%%%%%%%%%%%%%%%%%%%%%%%%%%%%%%

%%%%%%%%%%%%%%%%%%%%%%%%%%%%%%% FIG 5 %%%%%%%%%%%%%%%%%%%%
\begin{figure}[htb]
{\includegraphics[clip=true,width=7.cm, keepaspectratio,angle=0.]
{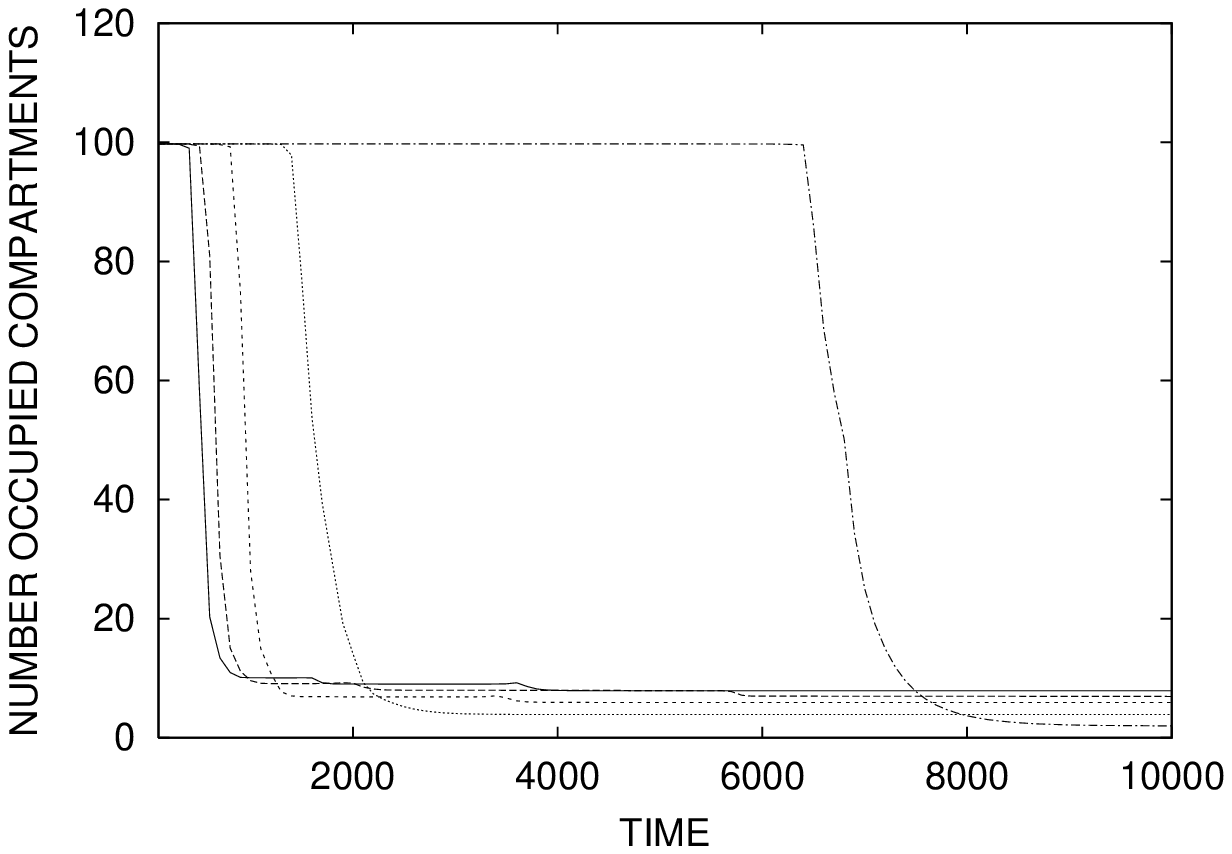}}
{\includegraphics[clip=true,width=7.cm, keepaspectratio,angle=0.]
{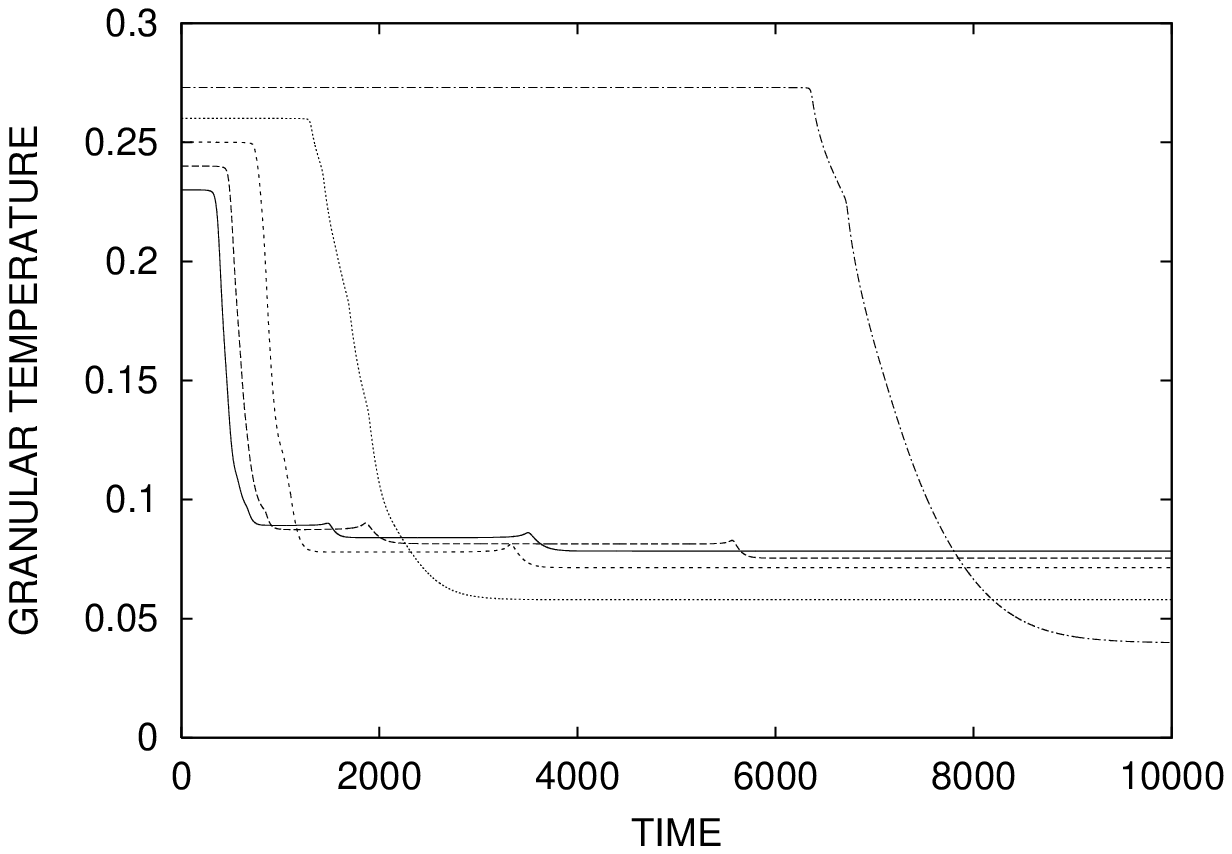}}
\caption
{Time evolution of the average number of occupied compartments, $f(t)$, 
relative to a system with
$M=100$ compartments and $180$ particles per compartment at $t=0$.
The curves refer from right to left to granular temperatures
$T_g=0.273,0.26,0.25,0.24,0.23$ (Fig. 5a).
The remaining parameters are those of fig~\ref{dispersion}.
In Fig. 5b we display the variation of the
average granular temperature 
for the same system (Fig. 5b).}
\label{tgmedia}
\end{figure}

%%%%%%%%%%%%%%%%%%%%%%%%%%%%%%%%%%%%%%%%%%%%%%%%%%%%%%%%%%%%%

%%%%%%%%%%%%%%%%%%%%%%%%%%%%%%% FIG 6 %%%%%%%%%%%%%%%%%%%%
\begin{figure}[htb]
{\includegraphics[clip=true,width=6.cm, keepaspectratio,angle=0]
{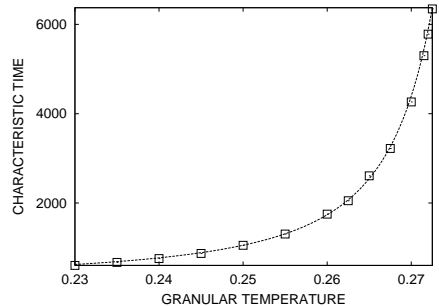}}
\caption
{Characteristic time associated with the position 
of the kink, shown in Fig. 5a, versus average granular temperature.
The system is a linear array with $M=100$
compartments. The remaining control parameters are 
the same as in fig~\ref{dispersion}. The behavior is
not Ahrrenius-like. 
The data can be fitted by a Vogel-Fulcher law 
$\tau_{vf}=A \exp[\Delta/(T_g-T_o)]$, where the 
fitting parameters are $A=200$, $\Delta=0.072$ and $T_o=0.293$.} 
\label{tau}
\end{figure}
%%%%%%%%%%%%%%%%%%%%%%%%%%%%%%%%%%%%%%%%%%%%%%%%%%%%%%%%%%%%%

In the present section we shall study numerically the properties of the
model. We have integrated the eqs. (\ref{eq:governing}) 
with $\alpha=0.7$, $V=100 \sigma^2$, $T_s=1$, $\tau_b=2$,
$\tau_s=0.5$. We have chosen $\sigma$ as unit length, $T_s$ as energy unit
and $2\tau_s$ as time unit.
The value of $T_b$ and of $\cal{N}$ and $M$ varied
from case to case, together with the initial conditions. 

The stability analysis discussed so far, describes only
the linear, early stage of the evolution. 
In order to explore the behavior in the late
non-linear regime we have solved numerically the governing equations
(\ref{eq:governing}). We considered a one dimensional
array of compartments, initially equally
populated, at the same granular
temperature, i.e. $N_i=N^{\ast}$ and $T_i=T^{\ast}$ and added a small
random perturbation.
According to the previous linear analysis,
two different behaviors can be observed as the control
parameters, such as the heat bath temperature and the 
average density, vary.
At fixed $N$, for large values of $T_b$, such that 
$T_g>T_c$, the initial perturbation 
is re-adsorbed diffusively,
while at small values of $T_b$ ($T_g<T_c$) the perturbation
is exponentially amplified.
In the latter case,
the collisional cooling determines a decrease of the local temperature
in correspondence of the regions more populated and clustering
begins. Some compartments, randomly
selected by the dynamics, act as germs for the nucleation process
illustrated in figs.~\ref{unstable}.
After the initial regime few compartments
grow at the expense of the remaining which become empty.
The distance between highly populated compartments increases,  
since they compete for particles. 
However, unlike the late stage
spinodal decomposition process \cite{spindecomp}, we observe that the 
domains do not grow in size,
but in height. This feature can be understood
because no saturation mechanism is present in the model, so that the
occupation number in a single
compartment can become of the order of the total population ${\cal N}$.
The growth process occurs by diffusion against the
density gradient, since particles move from low populated regions
toward highly populated regions. 
The evolution of the width, $W(t)$ (see 
fig. \ref{added})
\be
W(t)=\sqrt{\frac{1}{M}\sum_{i=1}^M [N_i^2(t)-(N^{\ast})^2]}
\ee
illustrates quantitatively how the process occurs. 
During the initial stage $W(t)$ remains much smaller than $N^{\ast}$
and only after a characteristic time $\tau$ it begins to rise
steeply. In fig.~\ref{added} we show $W(t)$ versus time and
the parametric plot $W(t)$ against the average value of the granular
temperature.  We notice that the latter plot recalls 
an order parameter versus temperature plot in a system undergoing
a phase transition.
%%%%%%%%%%%%%%%%%%%%%%%%%%%%%%%%%%%%%%%%%%%%%%

A second quantitative measure of the clustering
phenomenon is represented by the following statistical indicator
\be
h=-\sum_i^M \frac{N_i}{\cal N}\ln\left(\frac{N_i}{\cal N}\right)
\label{eq:entropy}
\ee 
The ``entropy'' $h$ is non negative, vanishes when all particles
are confined in a single compartment and takes on its
maximum value, $\ln(M)$, when all compartments are identically populated.
Thus $f=\exp(h)$ represents a measure of the number of occupied
compartments. In fig.~\ref{tgmedia} we display the evolution
of the average granular
temperature and of $f$ in various situations. 
Above $T_c$ the indicator $f$ relaxes
toward $M$, whereas in the low temperature region, due to clustering,
$f$ decreases toward a plateau value $P<M$.
Interestingly, such a relaxation time $\tau$ increases as the system approaches
the temperature $T_c$ from below. The temperature dependence of
$\tau$ close to $T_c$, displayed in fig. \ref{tau}, is 
consistent with the Vogel-Fulcher law 
\be
\tau_{vf}=A \exp[\Delta/(T-T_o)].
\label{VF}
\ee
The dependence of the characteristic time $\tau_{vf}$ on the temperature
is a direct consequence of eqs. (\ref{lambda}) and (\ref{a2}).
We also remark that the plateau value of $f$, reached by the system 
for $t>\tau_{vf}$ increases as $T_b$ decreases, indicating that
the system remains trapped in some metastable configurations.

We turn, now, attention to a different
process obtained by considering the evolution of
an initial
configuration, in which all
the particles are located inside a single compartment at $t=0$. 
According to the level of $T_b$
one can observe two different processes:
\noindent
a) for large $T_b$ the occupation number in the central cell
decays toward the fully symmetric state $N_i={\cal N}/M$
and $\lim_{t \to \infty} f(t)=M$;
\noindent
b) for small $T_b$ the occupation of the compartment remains constant.

\begin{figure}[htb]
\centerline{\includegraphics[clip=true,width=7.cm, keepaspectratio,angle=0]
{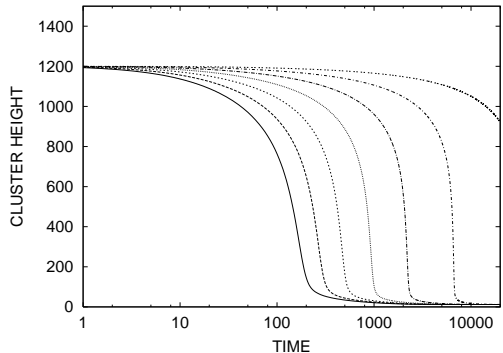}}
\caption
{Variation of the height of a cluster, initially
containing $1200$ particles, as a function of time for
temperatures
$T_b=0.30$, $0.35$, $0.40$, $0.45$, $0.50$, $0.55$ and $0.60$
from top to bottom.}
\label{piconeltempo}
\end{figure}

In fig.~\ref{piconeltempo} we display the variation in time of the population
in the compartment, where it was initially placed, for various
values of the heat bath temperature. We observe that $N(t)$ decreases
more and more slowly as the transition temperature is approached from above.
%%%%%%%%%%%%%%%%%%%%%%%%%%%%%%% FIG Cluster esplode %%%%%%%%%%%%%%%%%%%%

\begin{figure}[htb]
\centerline{\includegraphics[clip=true,width=7.cm, keepaspectratio,angle=0]
{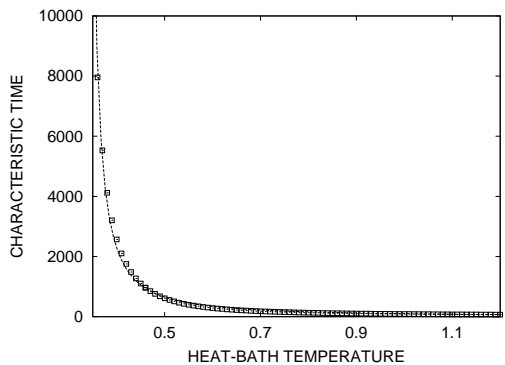}}
\caption
{Characteristic time associated with the explosion of a cluster of
$1200$ particles located at $t=0$ in a single compartment
versus the heath bath temperature.
The system consists of  
a linear array of $100$ compartments, while the remaining parameters are
the same as fig.~\ref{dispersion}}
\label{instabil}
\end{figure}

In fig.~\ref{instabil} is shown the
temperature dependence of the average time $\tau$ necessary
to wash-out the initial single cluster configuration. We observe 
that $\tau=C/(T_b-T_p)^{3/2}$ diverges at the crossover
temperature $T_p$. In particular the plot of the
occupation number versus time indicates the growth
of a plateau when $T_b \to T_p$
The smaller the temperature deviation from the limit of stability $T_p$ 
the longer the plateau. 

\begin{figure}[htb]
\centerline{\includegraphics[clip=true,width=5.cm, keepaspectratio,angle=-90]
{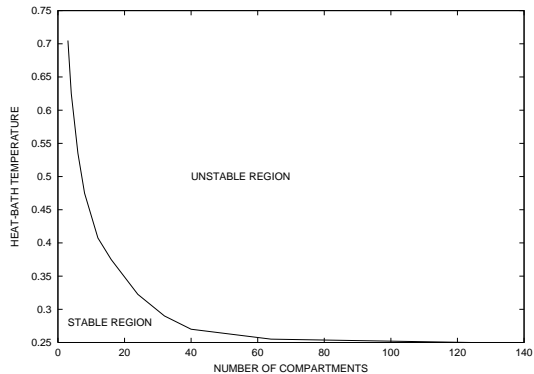}}
\caption
{Stability of an isolated cluster as a function of the
number of compartments. We show the
phase boundary between the stable and the unstable regime 
relative to a cluster
of $N=1200$ particles initially
placed in a single compartment. The vertical axis represents the
heat-bath temperature at which the transition
takes place, while the horizontal axis represents the number 
of compartments.}
\label{boundary}
\end{figure}

In fig.~\ref{boundary} we display the transition temperature at which 
a cluster of $N=1200$ particles placed in a cell at $t=0$ ``explodes'' as
a function of the number of boxes, $M$. Such an effect has been observed
experimentally by Lohse et al. 
\cite{Lohse} and explained in terms of their flux model.

%%%%%%%%%%%%%%%%%%%%%%%%%%%%%%% FIG 7 Cluster esplode %%%%%%%%%%%%%%%%%%%%

%%%%%%%%%%%%%%%%%%%%%%%%%%%%%%% FIG Cluster esplode %%%%%%%%%%%%%%%%%%%%

\begin{figure}[htb]
{\includegraphics[clip=true,width=6.cm, keepaspectratio,angle=0]
{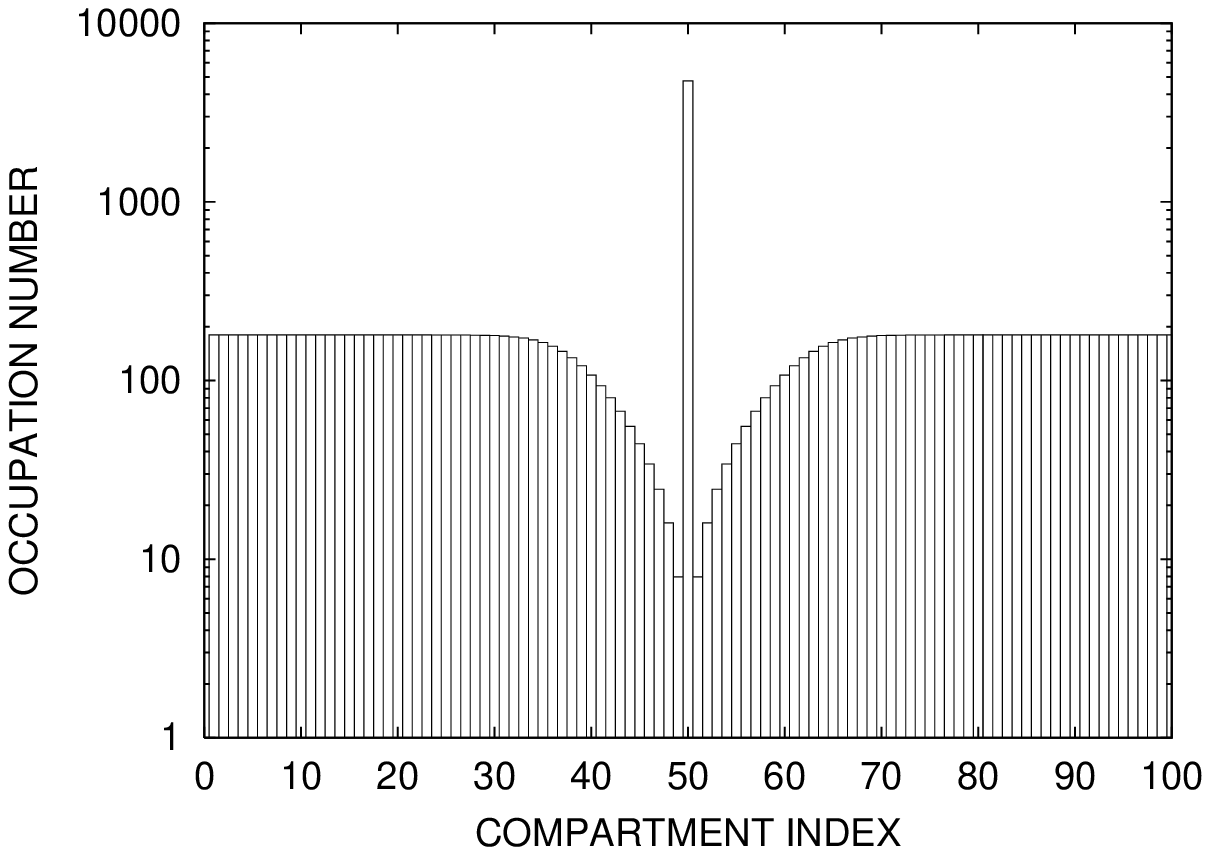}}
{\includegraphics[clip=true,width=6.cm, keepaspectratio,angle=0]
{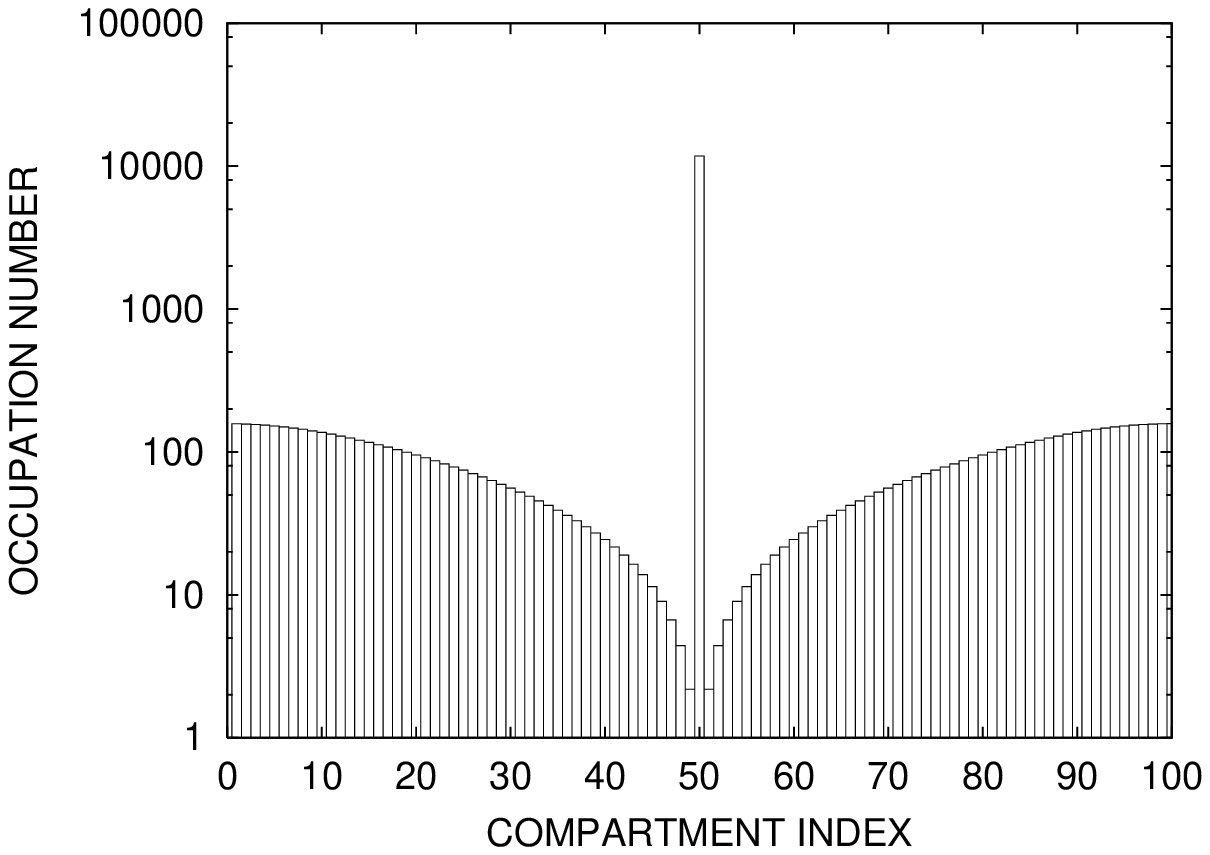}}
{\includegraphics[clip=true,width=6.cm, keepaspectratio,angle=0]
{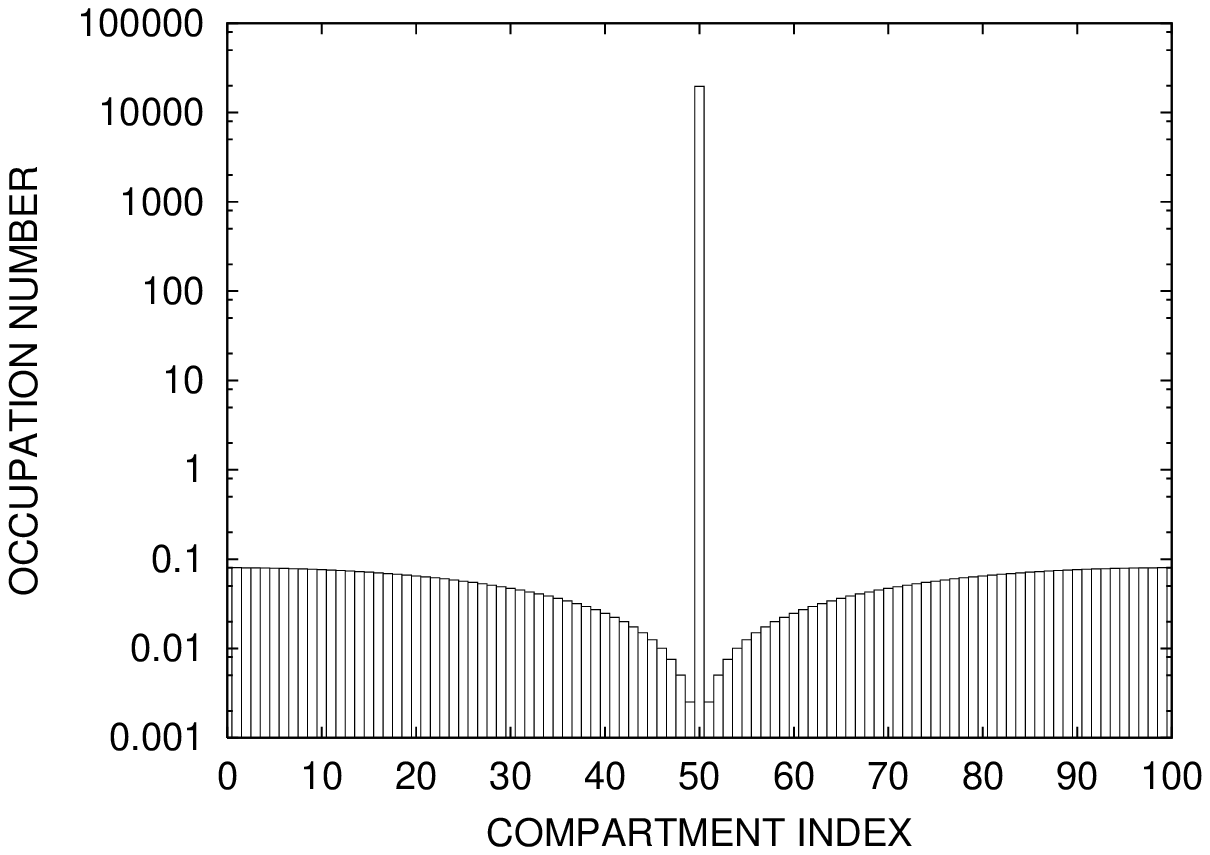}}
\caption
{Evolution of the depletion layer around a growing cluster at times 
$t=500$, $t=2500$ and $t=20000$, for $M=100$. The initial configuration is
$N_i=180$, $N_{50}=1800$ and $T_b=0.7$.
The system starts with a central
peak and a supersaturated ``bulk''. As the system evolves we observe the 
increasing thickness of the depletion layer. Matter accumulates in the central
compartment.}
\label{depleted}
\end{figure}

%%%%%%%%%%%%%%%%%%%%%%%%%%%%%%%%%%%%%%%%%%%%
Finally, we consider the inverse process, by which a cluster instead
of ``evaporating'', grows at the expenses of its neighbors.
Such a study also provides some information about 
the interfacial properties of the model.
We prepare the system in an
initial configuration where the population $N_c$ of a single
compartment is much larger than that of the remaining compartments.
If the profile were uniform with $N_i=N_{\infty}=180$ and $T_b=0.7$ the
state would be linearly stable with respect to perturbations. However,
if we place $N_c=1800$ particles in one of the compartments and
keep the remaining populations at $N_i=180$ we observe the 
following scenario:
the spike grows in height, while the compartments closer to it
slowly empty. As shown in fig. \ref{depleted}
the population profile develops a gradient
due to the flux of particles from the ``bulk'' to the ``tower''.
Its  shape is similar to the
depletion layer associated with the growth of a liquid droplet in a ``sea''
of over-saturated gas \cite{langer}. In addition, we observed that
since the flux terms in eq. (\ref{eq:gov1}) are very small
the local values of the granular temperature and 
of the occupation number are related by eq. (\ref{uniformT}).
 The profile varies almost linearly 
indicating that the process can be assimilated to a diffusion controlled
interfacial growth.

%%%%%%%%%%%%%%%%%%%%%%%%%%%%%%% FIG 9 Depletion layer %%%%%%%%%%%%%%%%%%%%

%\subsection{Revised model}
It is worthwhile to comment that
the model discussed above presents a shortcoming,
 since the local density can grow unbounded until all the particles
in the system occupy the same well. 
In fact, the approximation scheme does
not treat adequately the mutual repulsion 
between the particles at high densities. Particles are
allowed to pile up
and reach arbitrarily large packing fractions. In order to eliminate such a
shortcoming of the model, it is possible
to introduce phenomenologically a 
new ingredient into the theory.
The model was modified 
introducing an occupation number dependence in the characteristic 
time $\tau_s$:
\be
\tau_s(N_i)=\tau_o(1-N_i/M)^{1/2}
\ee
That is the probability
of escaping from a compartment becomes larger when it is occupied
by a large number of grains.
This correction is based on the results of microscopic studies \cite{fabiana}.
The form above has been chosen in order to preserve the form of the
equations as simple as possible and by 
phenomenological requirements. 
We observed that during the 
the early stages of the process, where the linear
analysis of section III applies,  there are no appreciable
changes to the situation described in the previous sections.
On the other hand, 
the late stage presents a different scenario.
When the compartments saturate the occupation number and the
granular temperature become locally stationary and the growth
proceeds in the neighboring compartments and so on.

%Let us notice that eq. (\ref{eq:gov1}) has the form postulated in the 
%flux models. Its right hand side represents the lattice Laplacian
%operator of a flux function $F_i=N_i e^{-T_s/T_i}$ which is a non monotonously
%increasing function of the occupation number.  
%This explains why the system possesses non uniform solutions. Upon adding
%a saturation term the flux function acquires a minimum.
%The shape of the flux function in this case becomes similar to that
%of a chemical potential in a two phase system.
\section{V. Conclusions}

To summarize, we introduced a model for compartmentalized
driven granular gases and studied it using the methods of 
kinetic theory. 
We have found a rather rich ``phase'' behavior and the emergence
of new qualitative properties as the number of particles
becomes sufficiently large.
We have pointed out that the system undergoes a long-wave length instability
and orders in a fashion similar to the process which occurs during
the spinodal decomposition in fluid mixtures. However, the late
stage of the process is radically different, because the granular
gas does not possess a surface tension mechanism
which restores homogeneity. Thus the usual competition
between bulk and surface free energy cost which determines the growth
of larger and larger domains is not at work here.

The present approach, in spite of its
simplicity, has the advantage with 
respect to the so called flux models of relating the microscopic
parameters to the macroscopic observables in a natural fashion. 
It can be extended
to treat granular mixtures, where the granular temperature of each
component must be treated as an independent variable to
be determined self-consistently \cite{Pagnani},\cite{Ioepuglio} . 
In addition,
the approach can be improved by including non Gaussian corrections
to the distribution function or by solving numerically by the Direct 
Monte Carlo simulation method the Boltzmann equation.
   
\section{Acknowledgments}

U.M.B.M. acknowledges the support of the 
Cofin MIUR ``Fisica Statistica di Sistemi Classici e Quantistici''.
He also thanks Giulio Costantini for reading the manuscript.

\end{document}